\documentstyle[11pt,aaspp4]{article}

\def\la{\mathrel{\hbox{\rlap{\hbox{\lower4pt\hbox{$\sim$}}}\hbox{$<$}}}}
\def\ga{\mathrel{\hbox{\rlap{\hbox{\lower4pt\hbox{$\sim$}}}\hbox{$>$}}}}

\slugcomment{Submitted to {\it The Astrophysical Journal (Letters)}}

\begin{document}

\title{Dust Sublimation and Fragmentation in the Circumburst Medium:  Evidence for a Large, Massive Cloud Origin for Gamma-Ray Bursts with Dark Optical Afterglows}

\author{Daniel E. Reichart\altaffilmark{1,2}}

\altaffiltext{1}{Department of Astronomy, California Institute of Technology, Mail Code 105-24, 1201 East California Boulevard, Pasadena, CA 91125}
\altaffiltext{2}{Hubble Fellow}

\begin{abstract}

The distances to which the optical flash destroys dust via sublimation, and the burst and afterglow change the size distribution of the dust via fragmentation, are functions of grain size.  Furthermore, the sublimation distance is a decreasing function of grain size, while the fragmentation distance is a decreasing function of grain size for large grains and an increasing function of grain size for small grains.  We investigate how these very different, but somewhat complementary, processes change the optical depth of the circumburst medium.  To this end, we adopt a canonical distribution of graphite and silicate grain sizes, and a simple fragmentation model, and we compute the post-burst/optical flash/afterglow optical depth of a circumburst cloud of constant density $n$ and size $R$ as a function of burst and afterglow isotropic-equivalent X-ray energy $E$ and spectral index $\alpha$, and optical flash isotropic-equivalent peak luminosity $L$:  This improves upon previous analyses that consider circumburst dust of a uniform grain size.  We find that circumburst clouds do not significantly extinguish ($\tau \la 0.3$) the optical afterglow if $R \la 10L_{49}^{1/2}$ pc, fairly independent of $n$, $E$, and $\alpha$, or if $N_H \la 5\times10^{20}$ cm$^{-2}$.  On the other hand, we find that circumburst clouds do significantly extinguish ($\tau \ga 3$) the optical afterglow if $R \ga 10L_{49}^{1/2}$ pc and $N_H \ga 5\times10^{21}$ cm$^{-2}$, creating a so-called `dark burst'.  The majority of bursts are dark, and as circumburst extinction is probably the primary cause of this, this implies that most dark bursts occur in clouds of this size and mass $M \ga 3\times10^5L_{49}$ M$_{\sun}$.  Clouds of this size and mass are typical of giant molecular clouds, and are active regions of star formation.

\end{abstract}


\section{Introduction}

Waxman \& Draine (2000) and Fruchter, Krolik \& Rhoads (2001) present mechanisms by which dust in the circumburst\footnote{By `circumburst', we mean within the cloud in which a gamma-ray burst occurs.} medium is destroyed by light from the optical flash (i.e., the reverse shock of the afterglow; e.g., Sari \& Piran 1999; Akerlof et al. 1999), and the size distribution of the dust is changed by X rays from the burst and early-time afterglow.  The former mechanism, sublimation, most strongly affects small grains:  The distance to which grains are sublimated is a decreasing function of grain size [$R_s(a) \sim a^b$, where $-1/2 \la b \la 0$].  The latter mechanism, fragmentation, most strongly affects grains of intermediate size:  The distance to which grains are fragmented is a decreasing function of grain size for large grains [$R_f(a) \rightarrow a^b$, where $-5/6 \la b \la -1/6$], and an increasing function of grain size for small grains [$R_f(a) \rightarrow a^{1/2}$].  In this letter, we investigate how the combination of these very different, but somewhat complementary, processes changes the optical depth of the circumburst medium.  To this end, we adopt a canonical distribution of graphite and silicate grain sizes, and a simple fragmentation model, and we compute the post-burst/optical flash/afterglow optical depth of a circumburst cloud of constant density $n$ and size $R$ as a function of burst and afterglow isotropic-equivalent X-ray energy $E$ and spectral index $\alpha$, and optical flash isotropic-equivalent peak luminosity $L$ (\S 2).  This improves upon the analyses of Waxman \& Draine (2000) and Fruchter, Krolik \& Rhoads (2001), which consider circumburst dust of a uniform grain size.

The results of this computation are directly applicable to the problem of the rapidly-, well-localized gamma-ray bursts with undetected, or dark, optical afterglows, or `dark bursts' for short.  About $2/3$ (e.g., Fynbo et al. 2001; Lazzati, Covino \& Ghisellini 2001) of the rapidly-, well-localized bursts fall under this category.  A wide variety of solutions have been proposed:  failure to image deeply enough quickly enough, circumburst extinction, host galaxy extinction unrelated to the circumburst medium, Galactic extinction, a second class of long-duration bursts with afterglows that are described by a very different parameterization of the relativistic fireball model, a second class of long-duration bursts with afterglows that do not arise from relativistic fireballs, and the following high redshift effects:  Lyman limit absorption in the source frame, absorption by the Ly$\alpha$ forest, absorption by excited molecular hydrogen in the circumburst medium, and source-frame extinction by the FUV component of the extinction curve (e.g., Fynbo et al. 2001; Lazzati, Covino \& Ghisellini 2001; Ramirez-Ruiz, Trentham \& Blain 2001; Reichart \& Yost 2001).  In Reichart \& Yost (2001), we show that although many of these explanations might contribute to some degree, circumburst extinction appears to be responsible for most of the dark bursts.  Taking this to be the case, we place constraints on the size and mass of the circumburst clouds of dark bursts in \S 3.  We draw conclusions in \S 4.

\section{Dust Sublimation and Fragmentation in the Circumburst Medium}

Dust sublimation occurs when a grain absorbs energy faster than it can thermally reradiate that energy.  When this happens, the grain heats up to a temperature of $T \approx 2300$ K, and the excess energy goes into breaking the bonds that hold atoms to the surface of the grain (e.g., Waxman \& Draine 2000).  If the energy flux on the grain is great enough long enough, the grain is sublimated away to nothing.  Grains in the path of an optical flash are destroyed almost instantaneously out to a distance 
\begin{equation}
R_s(a) \approx 12\left\{L_{49}e^{-\tau[R_s(a)]}Q(a)(a_3^{-1}+0.1)\right\}^{1/2}\,{\rm pc},
\label{Rs}
\end{equation}
where $a = 10^3a_3$ $\rm{\AA}$ is grain radius, $L = 10^{49}L_{49}$ erg s$^{-1}$ is the 1 -- 7.5 eV isotropic-equivalent peak luminosity of the optical flash, $\tau[R_s(a)]$ is the 1 -- 7.5 eV optical depth of the grains that are too large for the optical flash to sublimate, out to the sublimation distance, and $Q(a)$ is the efficiency at which grains absorb 1 -- 7.5 eV light (e.g., Waxman \& Draine 2000).  Hence, for $L \approx 10^{49}$ erg s$^{-1}$ (the 1 -- 7.5 eV isotropic-equivalent peak luminosity of the optical flash of GRB 990123; Waxman \& Draine 2000), an optically thin circumburst medium, and $Q(a) \sim 0.5$ (e.g., Mathis, Rumpl \& Nordsieck 1977; White 1979), the largest grains ($a_{max} \sim 10^4$ $\rm{\AA}$; e.g., Mathis, Rumpl \& Nordsieck 1977) are destroyed out to a few parsecs, and the smallest grains ($a_{min} \sim 50$ $\rm{\AA}$; e.g., Mathis, Rumpl \& Nordsieck 1977) are destroyed out to tens of parsecs.  Beyond these distances, the sublimation timescale is significantly longer than the duration of the optical flash, and consequently grains beyond these distances are relatively unaffected by sublimation (e.g., Waxman \& Draine 2000).\footnote{Fruchter, Krolik \& Rhoads (2001) show that X rays from the burst also sublimate grains, and that these X rays sublimate the largest grains to somewhat greater distances than does light from the optical flash.  However, since these X rays also fragment the grains at these distances to sufficiently small sizes to be sublimated by light from the optical flash (see below), we do not consider this mechanism in this letter.}

The dominant dust fragmentation mechanism is grain fission.  Grain fission occurs when a grain has been photoionized in excess of the maximum tensile strength of the grain's material, causing a Coulomb explosion that results in smaller grains (e.g., Waxman \& Draine 2000; Fruchter, Krolik \& Rhoads 2001).  Grains in the path of a burst and afterglow are fissioned on the emission timescale $\Delta t_X$ of the X rays at a distance  
\begin{equation}
R_f(a) \approx 120(\sigma_{-19}n_{23})^{1/2}x_K^{3/2}S_{10}^{-1/4}E_{51}^{1/2}4^{-\alpha/2}\left(1+\frac{\alpha}{3}\right)^{-1/2}a_3^{1/2}\left(a_3^{2/3}+\frac{x_K}{4}\right)^{-\frac{3+\alpha}{2}}\,{\rm pc},
\label{Rf}
\end{equation}
where $\sigma_{-19}n_{23} \approx 17$ and $x_K \approx 0.28$ for graphite, $\sigma_{-19}n_{23} \approx 2$ and $x_K \approx 1.9$ for silicate, $S = 10^{10}S_{10}$ dyn cm$^{-2}$ is the maximum tensile strength, $E = (\epsilon dE/d\epsilon)_{\epsilon = 1\,{\rm keV}} = 10^{51}E_{51}$ erg is the effective isotropic-equivalent energy of the burst and afterglow at 1 keV, and $\alpha$ is the energy spectral index (Fruchter, Krolik \& Rhoads 2001), and on a timescale 
\begin{equation}
\Delta t_f(a) = \Delta t_{X}[r/R_f(a)]^2
\label{Rf2}
\end{equation}
at other distances $r < R_f(a)$.  Consequently, grains at distances $r \ll R_f(a)$ fission multiple times (see below).  Hence, for $S_{10} \approx 1$ (e.g., Waxman \& Draine 2000; Fruchter, Krolik \& Rhoads 2001), $E_{51} \approx 1$, and $\alpha \approx 0$ (e.g., Frontera et al. 2000), the largest grains are fragmented out to tens of parsecs, and the smallest grains are fragmented out to hundreds of parsecs.  Consequently, grains, regardless of their size, are fragmented out to distances that are significantly greater than the distances to which they are destroyed by sublimation (Fruchter, Krolik \& Rhoads 2001).

We now adopt a canonical distribution of graphite and silicate grains, and determine how sublimation and fragmentation change this distribution as a function of grain size and distance.  For both grain species, we take their numbers to be distributed with grain size as $\sim a^{-7/2}$, but over different ranges:  $50 \la a \la 10^4$ $\rm{\AA}$ (graphite) and $250 \la a \la 2500$ $\rm{\AA}$ (silicate; e.g., Mathis, Rumpl \& Nordsieck 1977; Draine \& Lee 1984).  We mark these ranges with dashed lines in the top panels of Figure 1.  We take the circumburst medium to be constant in density.  

Since the optical flash is expected to be of equal or longer duration than the burst (e.g., Sari \& Piran 1999), we consider first the fragmentation of these grains, and second the sublimation of these grains.  The X-ray afterglow further fragments these grains on the timescale of the start time of the afterglow, but as we find that fragmenting these grains first vs. sublimating these grains first makes little difference, we do not consider a separate round of fragmentation in this letter.  We mark the fragmentation distance $R_f(a)$ (for $S_{10} = E_{51} = 1$ and $\alpha = 0$) with solid curves in the top panels of Figure 1.  Grains to the left of these curves, and between the dashed lines, are fragmented by the burst and afterglow.  In \S A, we present a simple fragmentation model in which grains are fissioned, often repeatedly, into two equal parts, which we approximate to be spherical.  Adopting this model, we show that the fragmented grains occupy the dotted regions in the top panels of Figure 1.  Notice that the grains closest to the burst have been fissioned repeatedly, and consequently have significantly smaller sizes than the grains at the fragmentation distance, which have been fissioned only once.  We have replotted the dotted regions, and the unchanged portions of the dashed regions, in the bottom panels of Figure 1.

We now consider the sublimation of grains within the dotted and dashed regions of the bottom panels of Figure 1.  Equation (\ref{Rs}) must be solved for the sublimation distance $R_s(a)$ numerically.  In Equation (\ref{Rs}), the 1 -- 7.5 eV optical depth of the grains that are too large to be destroyed by sublimation, out to a distance $r$, is given by 
\begin{equation}
\tau(r) = \sum_{i=1}^{2}\int_0^{r}\int_{a_s(r')}^{\infty}\pi a^2Q(a)n_i(a,r')dadr',
\label{tau}
\end{equation}
where $i = 1$ denotes graphite, $i = 2$ denotes silicate, $a_s(r)$ is the solution of Equation (\ref{Rs}) for $a$, and $n_i(a,r)$ is the number density of the grains of species $i$, which is related to the hydrogen density $n_H$ of the circumburst medium as described in Draine \& Lee (1984) and \S A.  Since the efficiency at which light of wavelength $\lambda$ is absorbed by small ($a \la 0.1\lambda$) grains scales with grain size (e.g., Draine \& Lee 1984), we take $Q(a) = a/(450$ ${\rm \AA})$ if $a < 450 $ ${\rm \AA}$, and 1 otherwise, where 450 ${\rm \AA}$ is one tenth of the log-central wavelength of the 1 -- 7.5 eV band:  This has the effect of canceling the dependence of $R_s(a)$ on $a$ when $a \la 450$ ${\rm \AA}$. 
Clearly, the solution for $R_s(a)$ depends on $n_H$:  From right to left in the bottom panels of Figure 1, we plot this solution (for $L_{49} = 1$) for $n_H \le 1$ cm$^{-3}$, $n_H = 10^3$ cm$^{-3}$, and $n_H \ge 10^5$ cm$^{-3}$.  Grains to the left of these curves are destroyed by sublimation.  Notice that as we increase the density of the circumburst medium to densities that are typical of dense clouds, the grains that are too large to be destroyed by sublimation increasingly extinguish the light from the optical flash, and consequently decrease the sublimation distance.  We have neglected the extinction caused by the sublimating grains, since Waxman \& Draine (2000) show that this is not a large effect.

In solving Equation (\ref{Rs}) for $R_s(a)$, one simultaneously solves Equation (\ref{tau}) for the differential optical depth $d\tau(r)/dr$ of the post-fragmentation/sublimation circumburst medium, which we plot in Figure 2 relative to that of the pre-fragmentation/sublimation circumburst medium for various values of $n_H$, $E$, $\alpha$, and $L$.  We find that the burst, optical flash, and afterglow burn through $\approx 10L_{49}^{1/2}$ pc of optical depth, fairly independent of $n_H$, $E$, $\alpha$, and $S$ (Equation {\ref{Rf}} depends on $S$ more weakly than it depends on $E$).  However, it is interesting to note that the optical depth beyond this distance is actually greater than its original value, because fragmentation increases the total grain cross section.\footnote{Consider a grain of initial size $a_0$ that is fragmented into $n = (a_0/a)^3$ grains of size $a$.  The total grain cross section increases by a factor of $n(a/a_0)^2[Q(a)/Q(a_0)]$, which $= a_0/a$ if $a > 450$ $\rm{\AA}$, $= a_0/(450\,\rm{\AA})$ if $a_0 > 450$ $\rm{\AA}$ and $a < 450$ $\rm{\AA}$, and $= 1$ if $a_0 < 450$ $\rm{\AA}$.}  We consider the cumulative optical depth, and implications for dark bursts, in \S 3.  

\section{Evidence for a Large, Massive Cloud Origin for Dark Bursts}

Neglecting for a moment extinction exterior to the circumburst cloud, we now consider the optical depth to a burst that is embedded a distance $r$ within its circumburst cloud, along the line of sight.  Prior to sublimation and fragmentation, the column to the burst is optically thin, which we take to mean $\tau \la 0.3$, if the hydrogen column density $N_H \la 5\times10^{20}$ cm$^{-2}$, and optically thick, which we take to mean $\tau \ga 3$, if $N_H \ga 5\times10^{21}$ cm$^{-2}$.  We mark column densities with dotted lines in Figure 3.  Since sublimation and fragmentation burn through $\approx 10L_{49}^{1/2}$ pc of optical depth (\S 2), the post-sublimation/fragmentation column to the burst is optically thin if $r$ is less than this distance or $N_H \la 5\times10^{20}$ cm$^{-2}$, and optically thick if $r$ is greater than this distance and $N_H \ga 5\times10^{21}$ cm$^{-2}$.  We show this in Figure 3 by plotting $\tau = 0.3$ and 3 for $L_{49} = E_{51} = 0.1$ (thin curves, left), 1 (thick curves), and 10 (thin curves, right).  Again, these results are fairly independent of the values of $E$, $\alpha$, and $S$ (\S 2).  

Consequently, modulo the value of $L$, bursts that occur to the lower left of the $\tau = 0.3$ curve have either relatively unextinguished optical afterglows, or optical afterglows that are extinguished by dust elsewhere in the host galaxy, or in our galaxy.  Bursts that occur to the upper right of the $\tau = 3$ curve have highly extinguished afterglows.  In Reichart \& Yost (2001), we show that circumburst extinction appears to be responsible for most of the dark bursts, and that Galactic extinction and host galaxy extinction unrelated to the circumburst medium account for no more than perhaps a few of the dark bursts detected to date.  Consequently, we find that most dark bursts have $r \ga 10L_{49}^{1/2}$ pc and $N_H \ga 5\times10^{21}$ cm$^{-2}$. 

We now estimate the sizes and masses of the circumburst clouds of the dark bursts.  Taking the clouds to be spherical, and taking the bursts to be located at the cloud centers, both of which are reasonable approximations on average, we find that most dark bursts occur in clouds of radius $R \ga 10L_{49}^{1/2}$ pc and mass $M \ga 3\times10^5L_{49}$ M$_{\sun}$ (Figure 4).  Clouds of this size and mass are typical of giant molecular clouds (e.g., Solomon et al. 1987), and are active regions of star formation.
  
\section{Discussion and Conclusions}

Waxman \& Draine (2000) introduced ten parsecs as the canonical distance to which the optical flash sublimates dust of a canonical size ($a = 10^3$ ${\rm \AA}$).  However, the sublimation distance is a function of grain size, and grain sizes span at least two orders of magnitude.  Furthermore, the fragmentation distance is also a function of grain size, with a different dependence on grain size, and we show in \S 2 that fragmentation actually increases the optical depth of the circumburst medium.  However, despite these complications, we again find ten parsecs to be the canonical distance for dust destruction in the circumburst medium, good perhaps to a factor of two, and a dependence on the isotropic-equivalent peak luminosity of the optical flash (\S 2).

Applying this result to the finding of Reichart \& Yost (2001) that circumburst extinction appears to be responsible for most of the dark bursts, we show in \S 3 that most of the dark bursts occur in clouds of size $R \ga 10L_{49}^{1/2}$ pc and mass $M \ga 3\times10^5L_{49}$ M$_{\sun}$.  Clouds of this size and mass are typical of giant molecular clouds, and are active regions of star formation.  This suggests that the dark bursts are the result of massive star death, and not neutron star coalescence, which would occur away from the stars' birth site.

In Reichart \& Yost (2001), we introduced a simple model in which the collimation angle of the burst's ejecta, and the column density to the burst through the circumburst cloud, determine whether a burst is dark or not:  Frail et al. (2001) show that the bursts for which redshifts have been measured draw upon a fairly standard energy reservoir of $\sim 3\times10^{51}$ erg (for an efficiency at which this energy is converted to gamma rays of $\eta \sim 0.2$; e.g., Beloborodov 2000), and that the wide range of isotropic-equivalent energies that have been implied for these bursts, from $\la 3\times10^{52}$ erg to $\ga 3\times10^{54}$ erg, is primarily the result of a wide range of collimation angles, with half angles ranging from $\la 0.05$ rad to $\ga 0.5$ rad.  If this is indeed the case, one expects a wide range of isotropic-equivalent peak luminosities for the optical flashes of these bursts, and consequently a wide range of optical depth burn distances, ranging from parsecs to several tens of parsecs.  Consequently, since most clouds, including most giant molecular clouds, tend to be less than several tens of parsecs across (e.g., Solomon et al. 1987), strongly collimated bursts likely burn completely through their circumburst clouds, while weakly collimated bursts likely often do not.\footnote{If the circumburst cloud is the central region of an ultraluminous infrared galaxy, as has been proposed by Ramirez-Ruiz, Trentham \& Blain (2001) to explain the dark bursts, even strongly collimated bursts would not burn through the hundreds of parsecs of optical depth that are typical of such regions (e.g., Solomon et al. 1997).  
However, we show in Reichart \& Price (2001) that the limited information that is available on the column densities, measured from absorption of the X-ray afterglow, of the dark bursts in not consistent with this idea.}

Consequently, in this simple model, the strongly collimated bursts have detectable optical afterglows regardless of the column density to the burst through the circumburst cloud, and the weakly collimated bursts have detectable optical afterglows if this column density is sufficiently low, and are dark if this column density is sufficiently high.  This simple model is consistent with the finding of Galama \& Wijers (2001) that bursts with detected optical afterglows occur in clouds of sizes and masses that are similar to what we find for the dark bursts, where their mass estimate is based on column densities measured from spectra of X-ray afterglows.  We verify and improve upon this finding in Reichart \& Price (2001).

\acknowledgements

Support for this work was provided by NASA through the Hubble Fellowship grant \#HST-SF-01133.01-A from the Space Telescope Science Institute, which is operated by the Association of Universities for Research in Astronomy, Inc., under NASA contract NAS5-26555.  I am also grateful to Don Lamb for the many discussions about dust in the circumburst medium that we have had over the years, and to Paul Price and Sarah Yost for critical readings.

\appendix

\section{Grain Fission Histories}

We now derive the boundaries of the dotted regions in Figure 1, and the number distribution of fragmented grains within these regions.  Consider a grain of initial size $a_0$ and initial photoionization cross section $\sigma_0 \sim a_0^3(a_0^{2/3}+x_K/4)^{-3-\alpha}$ (Fruchter, Krolik \& Rhoads 2001) at a distance $r$.  Let $t_1$ be the time it takes for the burst and afterglow to fission this grain at this distance.  From Equation (\ref{Rf}), $a_0(a_0^{2/3}+x_K/4)^{-3-\alpha}E = a_0^{-2}\sigma_0(d^2E/d\sigma dt)\sigma_0t_1 \sim (\sigma_0/a_0)^2t_1 =$ constant, so $t_1 \sim (a_0/\sigma_0)^{-2}$.  Modeling the fissioned grains as spheres, their sizes are $a_1 \approx 2^{-1/3}a_0 = 0.8a_0$.  Let $t_2$ be the time it takes for the burst and afterglow to fission the fissioned grains.  Since the fissioned grains have already been partially photoionized, $a_1^{-2}\sigma_1(0.5\sigma_0t_1 + \sigma_1t_2) =$ constant, or $0.8^{-2}(\sigma_1/\sigma_0)[0.5t_1 + (\sigma_1/\sigma_0)t_2] = t_1$, or $t_2/t_1 = 0.8^2(\sigma_1/\sigma_0)^{-2}-0.5(\sigma_1/\sigma_0)^{-1}$.  Let $t_3$ be the time it takes for the burst and afterglow to fission the grains again.  Then $0.8^{-4}(\sigma_2/\sigma_0)[0.5^2t_1 + 0.5(\sigma_1/\sigma_0)t_2 + (\sigma_2/\sigma_0)t_3] = t_1$, or $t_3/t_1 = 0.8^4(\sigma_2/\sigma_0)^{-2}-0.8^20.5(\sigma_2/\sigma_0)^{-1}(\sigma_1/\sigma_0)^{-1}$.  From here, it is not difficult to confirm that $t_n/t_1 = 0.8^{2(n-1)}(\sigma_{n-1}/\sigma_0)^{-2}-0.8^{2(n-2)}0.5(\sigma_{n-1}/\sigma_0)^{-1}(\sigma_{n-2}/\sigma_0)^{-1}$ for $n > 1$, where 
\begin{equation}
\frac{\sigma_n}{\sigma_0} = 0.8^{3n}\frac{(0.8^na_0)^{2/3}+x_k/4}{a_0^{2/3}+x_k/4}
\end{equation}
(since $a_n = 0.8^na_0$), and $n = \log{(a_n/a_0)}/\log{0.8}$.

Consider now a grain of initial size $a_0$ at a distance $r$ that is fragmented into grains of size $a_n$ on the emission timescale $\Delta t_X$ of the X rays.  Then $\Delta t_X = t_1\sum_{i=1}^n(t_i/t_1)$. Using Equation (\ref{Rf2}), $t_1 =\Delta t_X[r/R_f(a_0)]^2$.  Hence, $r(a_n) = R_f(a_0)[\sum_{i=1}^n(t_i/t_1)]^{-1/2}$.  The boundaries of the dotted regions in Figure 1 are then given by setting $a_0 = a_{max}$ (top boundary), $a_0 = a_{min}$ (bottom boundary), and $a_0 = a_f(r)$ (right boundary), where $a_f(r)$ is the solution of Equation (\ref{Rf}) for $a$.  The number distribution of fragmented grains within these regions is $\propto (a_0/a)^3a_0^{-7/2}(da_0/da)$ (\S 2).

\clearpage

\figcaption[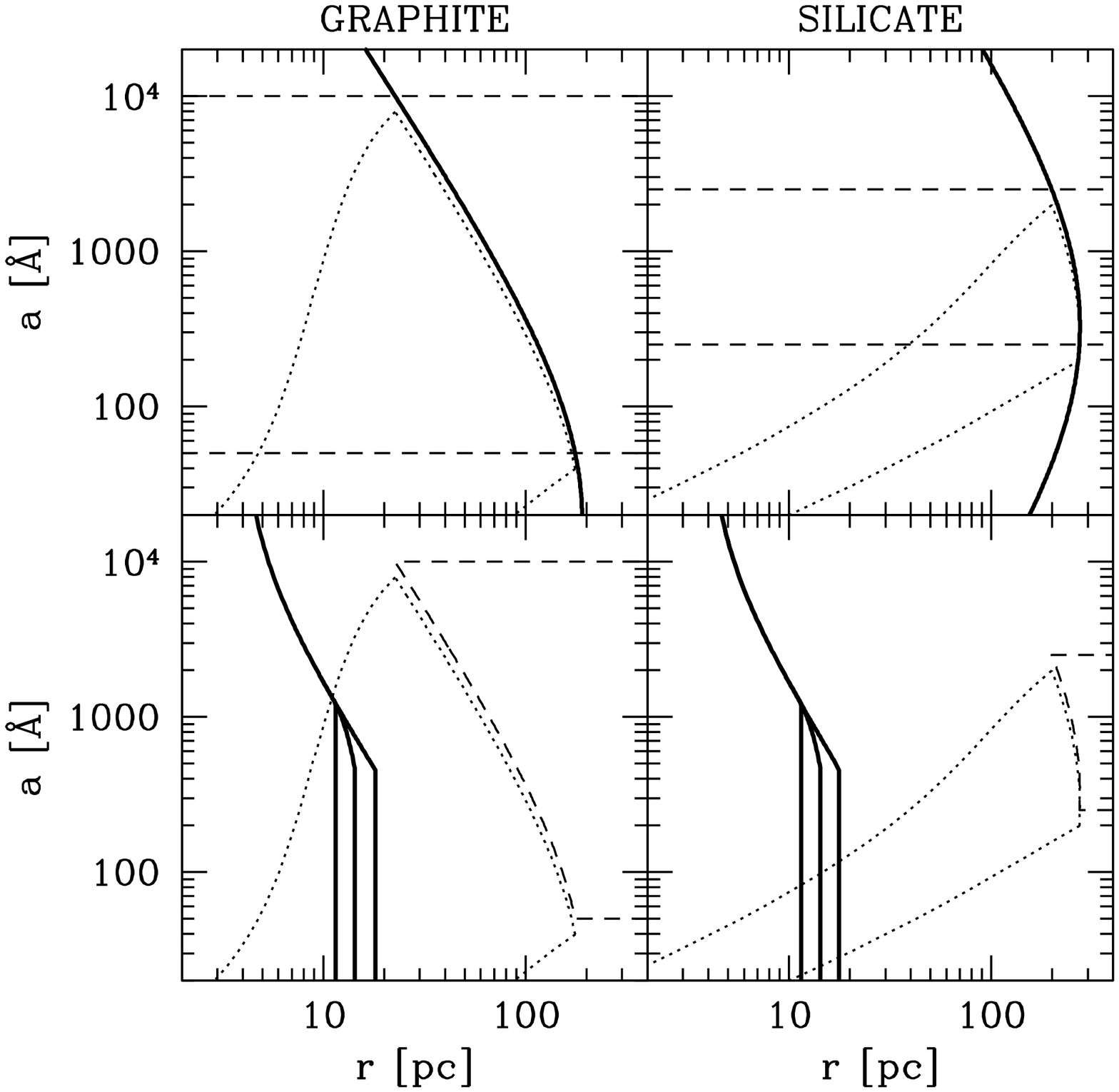]{Graphite (left panels) and silicate (right panels) grain size $a$ and distance $r$ distributions.  Top panels:  The dashed lines mark the boundaries of these distributions before fragmentation and sublimation.  The solid curves mark the distance to which the burst and afterglow fragment grains of size $a$ (for $S_{10} = E_{51} = 1$ and $\alpha = 0$).  The dotted curves mark the boundaries of the distributions of the fragmented grains.  Bottom panels:  The dashed curves mark the boundaries of the distributions of the grains that are too distant for the burst and afterglow to fragment, and the dotted curves are the same as in the top panels.  The solid curves mark the distance to which the optical flash sublimates grains of size $a$ (for $L_{49} = 1$) for various hydrogen densities of the circumburst medium, from right to left:  $n_H \le 1$ cm$^{-3}$, $n_H = 10^3$ cm$^{-3}$, and $n_H \ge 10^5$ cm$^{-3}$.  Only grains to the right of these curves survive (\S 2).\label{subfrag.eps}}

\figcaption[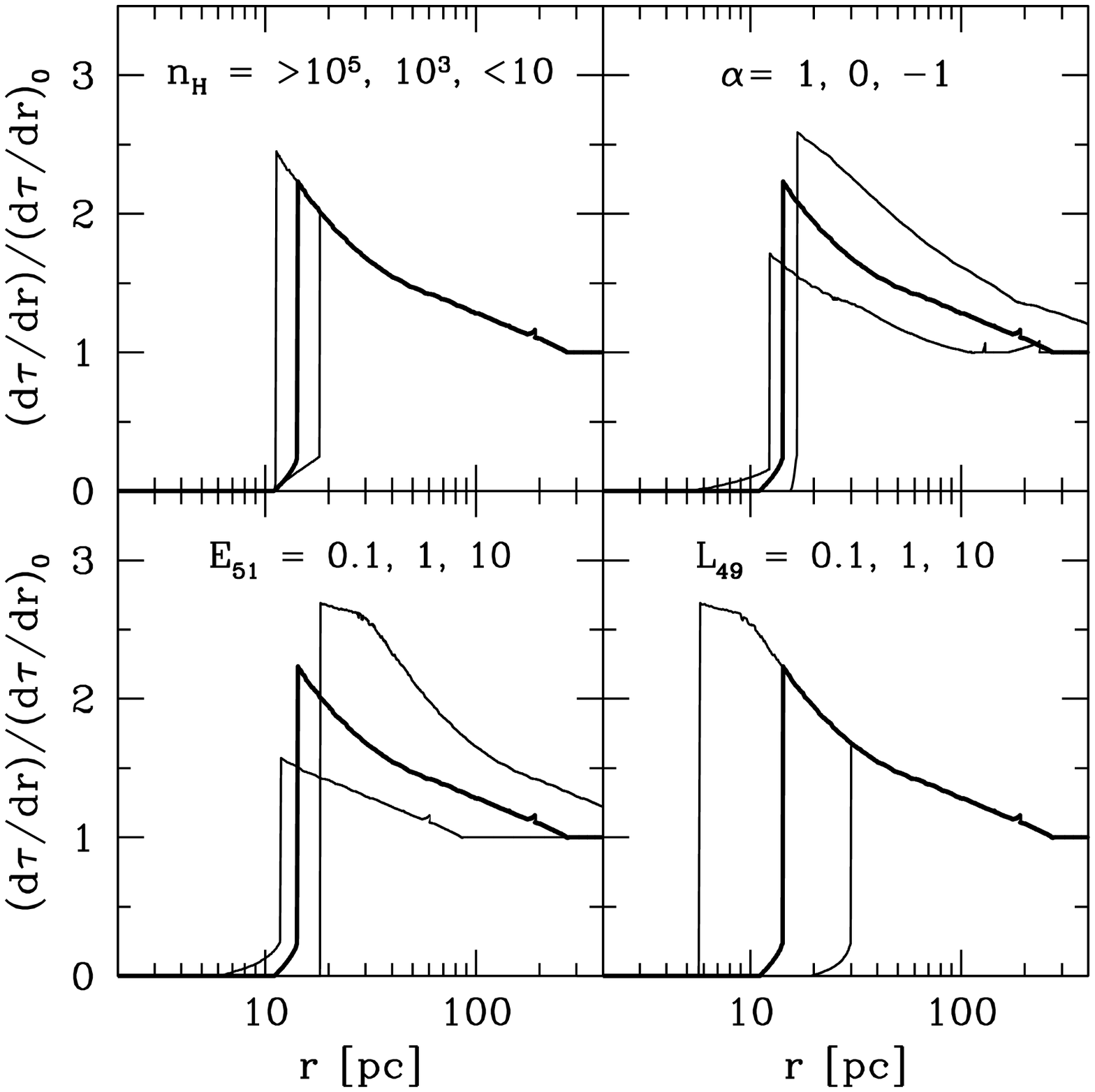]{Differential optical depth $d\tau(r)/dr$ of the circumburst medium after fragmentation and sublimation, relative to the original differential optical depth.  The thick curves are for $n_H = 10^3$ cm$^{-3}$, $\alpha = 0$, $E_{51} = L_{49} = S_{10} = 1$.  In the top left panel, we show the effect of increasing (left curve) and decreasing (right curve) the hydrogen density of the circumburst medium.  In the top right panel, we show the effect of increasing (left curve) and decreasing (right curve) the energy spectral index.  In the bottom left panel, we show the effect of decreasing (left curve) and increasing (right curve) the isotropic-equivalent X-ray energy of the burst and afterglow, and in the bottom right panel, we show the effect of decreasing (left curve) and increasing (right curve) the isotropic-equivalent peak luminosity of the optical flash (\S 2).\label{dtaudr.eps}}

\figcaption[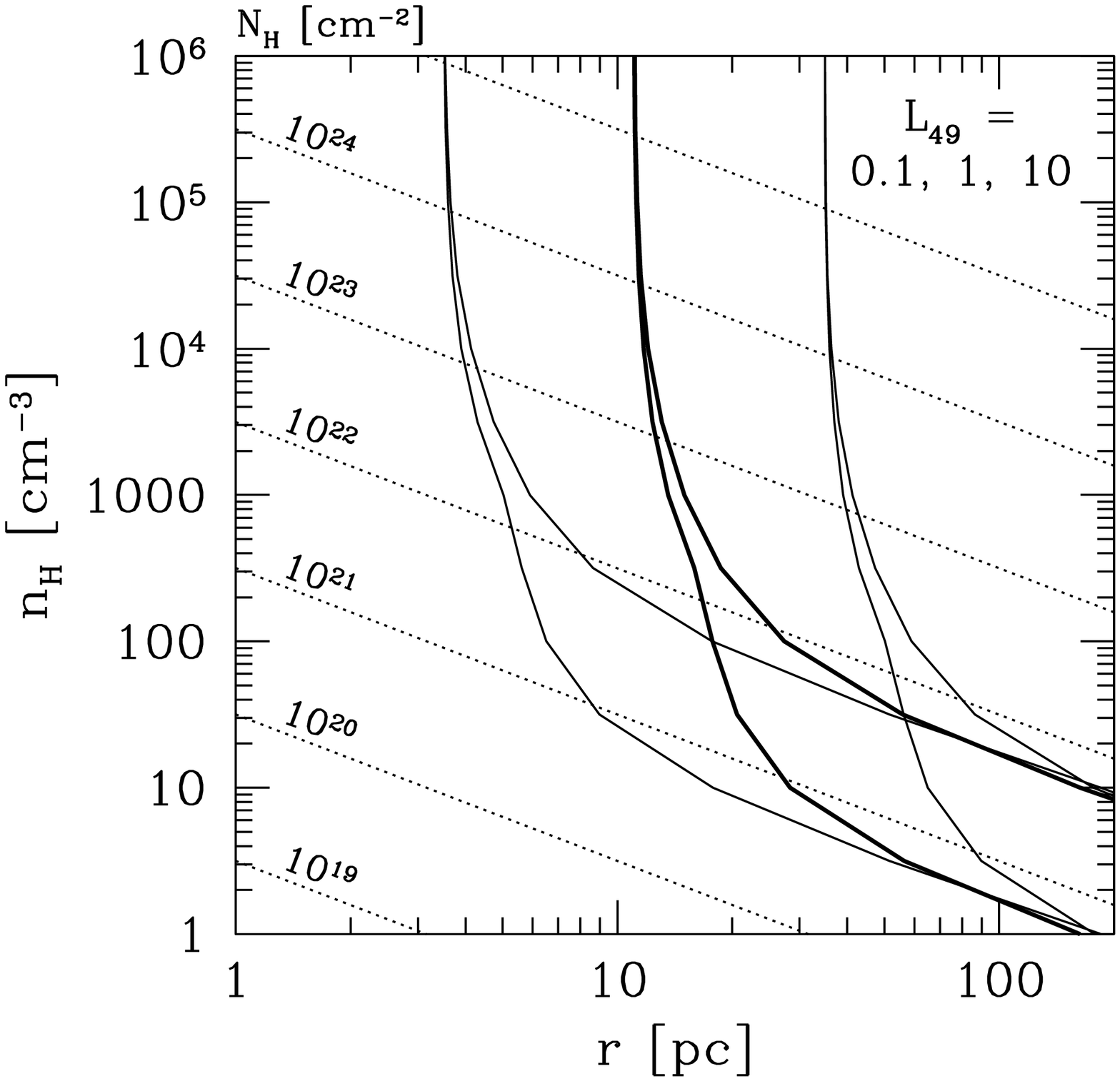]{Post-sublimation/fragmentation optical depth $\tau$ to a burst that is embedded a distance $r$, along the line of sight, within a cloud of constant hydrogen density $n_H$.  The three pairs of solid curves mark $\tau = 0.3$ (lower left) and 3 (upper right) for $L_{49} = 0.1$ (left), 1 (center), and 10 (right).  For a given $L$, bursts that occur to the lower left of the $\tau = 0.3$ curve have afterglows that are relatively unextinguished by the circumburst cloud, and bursts that occur to the upper right of the $\tau = 3$ curve have highly extinguished afterglows.  The dotted lines mark constant hydrogen column densities $N_H$ (\S 3).\label{taunh.eps}}

\figcaption[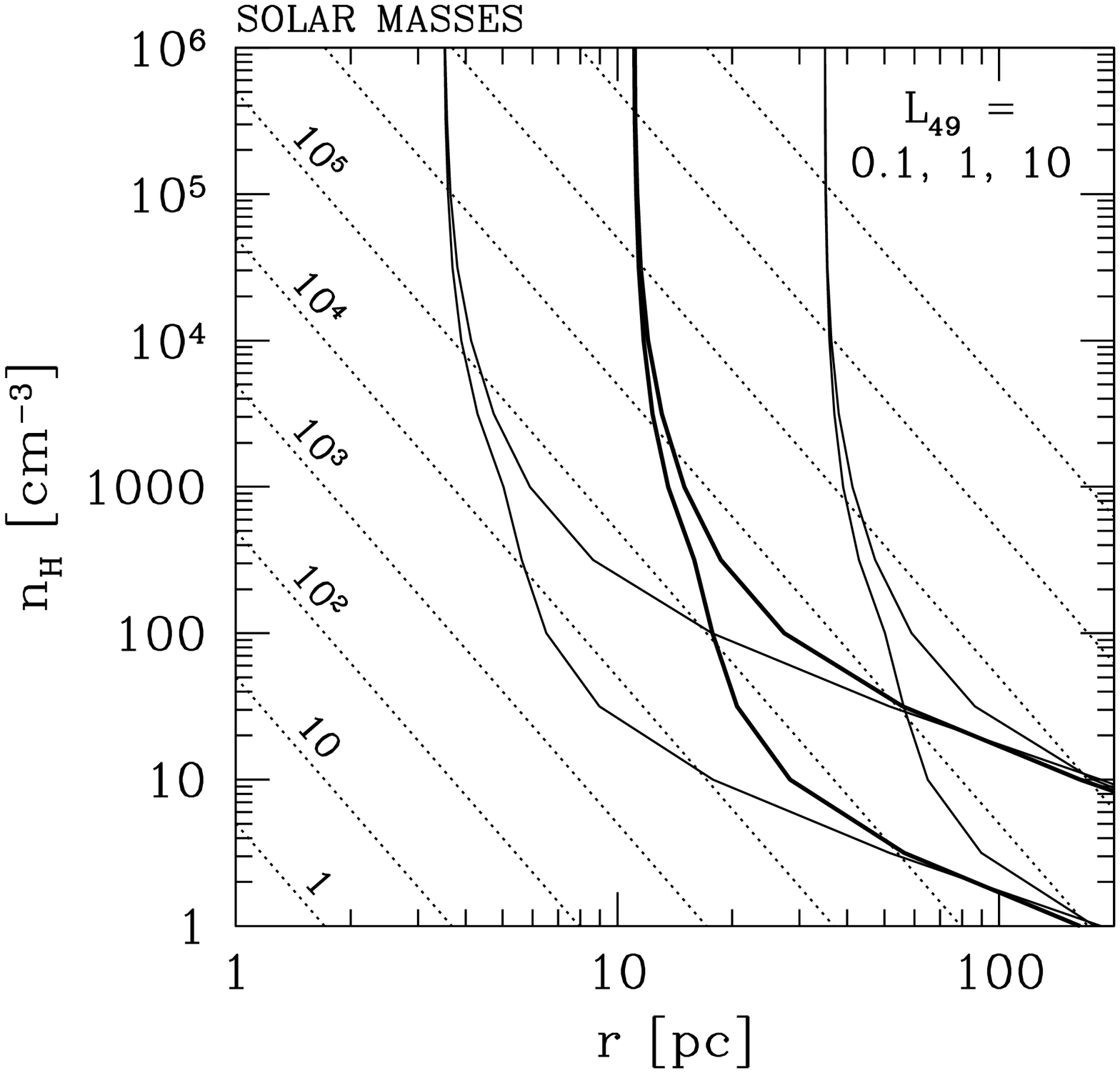]{Post-sublimation/fragmentation optical depth $\tau$ to a burst that is embedded a distance $r$, along the line of sight, within a cloud of constant hydrogen density $n_H$.  The three pairs of solid curves are the same as in Figure 3.  The dotted lines mark constant cloud masses $M$, where we have approximated the clouds to be spherical, and the bursts to be located at the cloud centers (\S 3).\label{taum.eps}}

\clearpage

\setcounter{figure}{0}

\begin{figure}[tb]
\plotone{subfrag.eps}
\end{figure}

\begin{figure}[tb]
\plotone{dtaudr.eps}
\end{figure}

\begin{figure}[tb]
\plotone{taunh.eps}
\end{figure}

\begin{figure}[tb]
\plotone{taum.eps}
\end{figure}

\end{document}